\documentclass[paper,notoc]{JHEP3}

\usepackage{epsfig}
\usepackage{graphicx}
\usepackage{feynmp}

\usepackage{amsmath}
\usepackage{amsfonts}
\usepackage{graphicx}

\def\beq{\begin{equation}}
\def\eeq{\end{equation}}
\newcommand{\bea}{\begin{eqnarray}}
\newcommand{\eea}{\end{eqnarray}}

\def\Eqn#1{Eq.~(\ref{#1})}

\def\eqns#1#2{Eqs.~(\ref{#1}) and~(\ref{#2})}
\def\eqnss#1#2{Eqs.~(\ref{#1})-(\ref{#2})}

\def\bsp#1\esp{\begin{split}#1\end{split}}

\newcommand{\eps}{\epsilon}

\newcommand{\ord}{\begin{cal}O\end{cal}}

\def\P{{\cal P} }

\def\N{{\cal N} }
\def\C{{\cal C} }

\newcommand{\rd}{\mathrm{d}}

\def\bit#1\eit{\begin{itemize}#1\end{itemize}}
\def\ben#1\een{\begin{enumerate}#1\end{enumerate}}

\newenvironment{sloppyequation}[0]{\sloppy\begin{flushleft}\hspace*{0.75cm}\(}{\)\end{flushleft}\fussy}
\newenvironment{sloppytext}[0]{\sloppy\begin{flushleft}}{\end{flushleft}\fussy}

\newcommand{\beqsloppy}{\begin{sloppyequation}}
\newcommand{\eeqsloppy}{\end{sloppyequation}}
\newcommand{\btxtsloppy}{\begin{sloppytext}}
\newcommand{\etxtsloppy}{\end{sloppytext}}

\title{An Analytic Result for the Two-Loop Hexagon Wilson Loop in $\N = 4$ SYM}

\author{Vittorio Del Duca\\
Istituto Nazionale di Fisica Nucleare\\
Laboratori Nazionali di Frascati\\
00044 Frascati (Roma), Italy\\
       E-mail: \email{delduca@lnf.infn.it}}

\author{Claude Duhr\\
Institute for Particle Physics Phenomenology, 
University of Durham\\ Durham, DH1 3LE, U.K.\\
E-mail: \email{claude.duhr@durham.ac.uk}}

\author{Vladimir A. Smirnov\\
Nuclear Physics Institute of Moscow State University\\
Moscow 119992, Russia\\
E-mail: \email{smirnov@theory.sinp.msu.ru}}

\abstract{In the planar $\begin{cal}N\end{cal}=4$ supersymmetric Yang-Mills theory,
the conformal symmetry constrains multi-loop $n$-edged Wilson loops to be basically
given in terms of the one-loop $n$-edged Wilson loop, augmented, for $n\ge 6$, 
by a function of conformally invariant cross ratios. We identify a class of kinematics for
which the Wilson loop exhibits exact Regge factorisation and which
leave invariant the analytic form of the multi-loop $n$-edged Wilson loop.
In those kinematics, the analytic result for the Wilson loop is the same as in general kinematics,
although the computation is remarkably simplified with respect to general kinematics. 
Using the simplest of those kinematics,
we have performed the first analytic computation of the 
two-loop six-edged Wilson loop in general kinematics.}

\preprint{IPPP/09/92, DCPT/09/184}

\begin{document}

\section{Introduction}
\label{sec:intro}

In the planar $\begin{cal}N\end{cal}=4$ supersymmetric Yang-Mills (SYM) theory,
Anastasiou, Bern, Dixon and Kosower (ABDK)~\cite{Anastasiou:2003kj} proposed an iterative 
structure for the colour-stripped two-loop scattering amplitude with an arbitrary number $n$ of external 
legs in a maximally-helicity violating (MHV) configuration. Writing at any loop order $L$, the amplitude
$M_n^{(L)}$ as the tree-level amplitude, $M_n^{(0)}$, which depends on the helicity configuration, 
times a scalar function, $m_n^{(L)}$,
\beq
M_n^{(L)} = M_n^{(0)}\ m_n^{(L)}\,,
\eeq
the proposed iteration formula for the two-loop MHV amplitude $m_n^{(2)}(\eps)$ was
\beq
m_n^{(2)}(\eps) = \frac{1}{2} \left[m_n^{(1)}(\eps)\right]^2
+ f^{(2)}(\eps)\, m_n^{(1)}(2\eps) + C^{(2)} + \ord(\eps)\, .\label{eq:ite2bds}
\eeq
Thus the two-loop amplitude is determined in terms of 
the one-loop MHV amplitude $m_n^{(1)}(\eps)$ evaluated through to $\ord(\eps^2)$
in the dimensional-regularisation parameter $\eps=(4-d)/2$,
the constant $C^{(2)} = -\zeta_2^2/2$, and the function $f^{(2)}(\eps) = -\zeta_2 - \zeta_3\eps - \zeta_4\eps^2$,
with $\zeta_i = \zeta(i)$ and $\zeta(z)$ the Riemann zeta function.
%

Subsequently, Bern, Dixon and one of the present authors (BDS) proposed an all-loop resummation 
formula~\cite{Bern:2005iz} for the colour-stripped $n$-point MHV amplitude, which implies a tower
of  iteration formulae, allowing one to determine the $n$-point amplitude at a given number of
loops in  terms of amplitudes with fewer loops, evaluated to higher orders of $\eps$. BDS checked
that the ansatz is correct for the three-loop four-point amplitude, by evaluating analytically 
$m_4^{(3)}(\eps)$ through to finite terms, as well as $m_4^{(2)}(\eps)$ through to $\ord(\eps^2)$ and
$m_4^{(1)}(\eps)$ through to $\ord(\eps^4)$. 
The ansatz has been proven to be correct also for the
two-loop five-point  amplitude~\cite{Bern:2006vw,Cachazo:2008vp}, for which $m_5^{(2)}(\eps)$ has
been computed numerically through to finite terms, as well as $m_5^{(1)}(\eps)$ through to $\ord(\eps^2)$.

Using the AdS/CFT correspondence, Alday and Maldacena showed that in the strong-coupling limit
the ansatz must break down for amplitudes with a large number of legs~\cite{Alday:2007he}. 
At weak coupling, the computation of the two-loop six-edged Wilson loop~\cite{Drummond:2007bm}
led to conclude that either the ansatz or the duality relation between amplitudes and Wilson loops
were to break down for two-loop six-point amplitudes.
Likewise, there were hints of a failure of the ansatz
from the six-point amplitude analysed in the multi-Regge kinematics in a Minkowski 
region~\cite{Bartels:2008ce,Bartels:2008sc, Schabinger:2009bb}.
The accumulating evidence against the ansatz provoked the numerical calculation of 
$m_6^{(2)}(\eps)$ through to finite terms and of $m_6^{(1)}(\eps)$ through to $\ord(\eps^2)$, by which
the ansatz was demonstrated to fail~\cite{Bern:2008ap}, and where
it was shown that the finite pieces of the parity-even part of $m_6^{(2)}(\eps)$
are incorrectly determined by the ansatz (although
the parity-odd part of $m_6^{(2)}(\eps)$ does satisfy the ansatz~\cite{Cachazo:2008hp}).
In particular, it was shown numerically that the two-loop remainder function,
defined as the difference between the two-loop amplitude and the ansatz for it,
\beq
R_n^{(2)} = m_n^{(2)}(\eps) - \frac{1}{2} \left[m_n^{(1)}(\eps)\right]^2
- f^{(2)}(\eps)\, m_n^{(1)}(2\eps) - C^{(2)} + \ord(\eps)\, ,\label{eq:discr}
\eeq
is different from zero for $n = 6$,
where $R_n^{(2)}$ is a function of the kinematical parameters of the $n$-point amplitude,
but a constant with respect to $\eps$. However, the analytic form of $R_6^{(2)}$ was not computed.

In the strong-coupling limit, Alday and Maldacena~\cite{Alday:2007hr} showed that 
planar scattering amplitudes exponentiate like in the ansatz, and 
suggested that the vacuum expectation value of the $n$-edged Wilson
loop could be related to the $n$-point amplitude in $\begin{cal}N\end{cal}=4$ SYM.
At weak coupling,
the agreement between the light-like Wilson loop and the (parity-even
part of the) MHV amplitude has been verified for the one-loop four-edged~\cite{Drummond:2007aua}
and $n$-edged~\cite{Brandhuber:2007yx} Wilson loops, and for the two-loop 
four-edged~\cite{Drummond:2007cf}, five-edged~\cite{Drummond:2007au} and 
six-edged~\cite{Drummond:2007bm,Drummond:2008aq} Wilson loops. 

Furthermore, it was shown that the $L$-loop light-like Wilson loop exhibits a conformal symmetry, 
and that the solution of the Ward identity for a special conformal boost is the ansatz, augmented,
for $n\ge 6$, by a function $R_{n,WL}^{(L)}$ of conformally invariant cross-ratios~\cite{Drummond:2007au}. 
Because of the duality between Wilson loops and amplitudes at one and two loops, the function of the 
conformally invariant cross-ratios, $R_{n,WL}^{(2)}$, can be identified as the remainder function of \Eqn{eq:discr}. 

In Refs.~\cite{Drummond:2008aq,Anastasiou:2009kna}, the two-loop $n$-edged Wilson loop 
has been given in terms of Feynman-parameter-like integrals. Furthermore, in Ref.~\cite{Anastasiou:2009kna}
a numerical algorithm has been set up, which is valid for the two-loop $n$-edged Wilson loop and
by which the seven-edged and eight-edged Wilson loops have been computed 
(although the corresponding MHV amplitudes are not known\footnote{The parity-even part of
the two-loop $n$-point MHV amplitude has been given in terms of integral functions, yet to be 
evaluated~\cite{Vergu:2009tu}.}). Thus, also the remainder
functions $R_{7,WL}^{(2)}$ and $R_{8,WL}^{(2)}$ of the Wilson loops are
known numerically, and the numerical evidence~\cite{Anastasiou:2009kna} confirms that they
are functions of conformally invariant cross-ratios only. However, their analytic form is in general 
unknown.

In fact, before this letter the remainder function had been computed analytically only in the strong coupling regime:
for the two-loop six-edged Wilson loop in the particular kinematic configuration in which the cross ratios
coincide~\cite{Alday:2009dv}, and otherwise in a 
particular kinematic set-up for which only $2n$-edged polygons are allowed~\cite{Alday:2009yn}. 
In that set-up, the simplest non-trivial remainder function is the one of the two-loop eight-edged
Wilson loop. In the weak coupling regime, and in the same kinematics, $R_{8,WL}^{(2)}$ has been computed
numerically in Ref.~\cite{Brandhuber:2009da}, where a numerical evidence has been found for a linear relation
between the remainder functions at weak and strong couplings.

In this letter, we give a brief account of the first analytic computation at weak coupling of the 
two-loop six-edged Wilson loop in general kinematics. The computation has been done in the Euclidean region
in $D=4-2\eps$ dimensions, where the result is real. 

In Sec.~\ref{sec:wl}, we write the two-loop Wilson loop in terms of the one-loop Wilson loop plus a
remainder function $R_{n,WL}^{(2)}$. For $n=6$, $R_{6,WL}^{(2)}$ is a function of the three conformally 
invariant cross ratios, $u_1,u_2,u_3$. It has been computed numerically 
in Refs.~\cite{Bern:2008ap,Drummond:2008aq,Anastasiou:2009kna} in arbitrary kinematics, 
constrained only by momentum conservation.
However, it suffices to compute $R_{6,WL}^{(2)}$ in any kinematical limit which 
does not modify the analytic dependence of $R_{6,WL}^{(2)}$ on $u_1,u_2,u_3$. To that end, we recall that
the $L$-loop four-edged Wilson loop is not modified by the Regge limit~\cite{Drummond:2007aua}.
In such a limit $w_4^{(L)}$ undergoes an exact Regge factorisation.
So is the case of the five-edged Wilson loop in the multi-Regge 
kinematics~\cite{Bartels:2008ce,Brower:2008nm}.
However, the six-edged Wilson loop is modified by the multi-Regge kinematics:
the three conformally invariant cross ratios are not invariant in such a limit~\cite{Brower:2008nm}.
Less constraining Regge limits have been analysed in Ref.~\cite{DelDuca:2008jg}.
The simplest of those limits to feature an exact Regge factorisation of $w_6^{(L)}$ 
is the quasi-multi-Regge kinematics (QMRK) of a pair along the ladder~\cite{Fadin:1989kf,DelDuca:1995ki}.

In Sec.~\ref{sec:qmrk}, we recall the QMRK of a pair along the ladder for the six-edged Wilson loop,
and we show that the QMRK of three-of-a-kind along the ladder~\cite{Del Duca:1999ha} for the seven-edged 
Wilson loop, the QMRK of four-of-a-kind along the ladder~\cite{claude} for the eight-edged Wilson 
loop, and in general the QMRK of a cluster of $(n-4)$-of-a-kind along the ladder for the $n$-edged 
Wilson loop  do not modify the analytic dependence of $w_n^{(L)}$ on the conformally invariant cross ratios.
That is, this class of kinematics exhibits an exact Regge factorisation of $w_n^{(L)}$. Thus, the result 
for $w_n^{(L)}$ in these kinematics is the same as the result in general kinematics, although 
the computation is remarkably simplified with respect to the same computation in general kinematics.
Finally, we note that 
although in Sec.~\ref{sec:w62} we apply the analysis of Sec.~\ref{sec:qmrk} to the computation of the 
six-edged two-loop Wilson loop, nothing of what we consider in Sec.~\ref{sec:qmrk} is specific to 
two loops: The analysis of Sec.~\ref{sec:qmrk} is valid for any number of loops.

In Sec.~\ref{sec:w62}, we brief on how the Feynman-parameter-like integrals of the two-loop
six-edged Wilson loop have been computed in the QMRK of a pair along the ladder, and on the type
of functions which appear in the final result. Because of the exact Regge factorisation, the ensuing remainder function 
is valid in general kinematics. 
It can be expressed as a linear combination of multiple polylogarithms of uniform transcendental weight four.
However, the result is far too long to be reported in this letter. We present it in an electronic form at {\tt www.arxiv.org} 
where a text file containing the {\tt Mathematica} expression for the remainder function is provided.

\section{The two-loop Wilson loop}
\label{sec:wl}

The Wilson loop is defined through the path-ordered exponential,
\beq
W[\C_n] = {\rm Tr}\ \P\ {\rm exp} \left[ ig \oint {\rm d}\tau \dot{x}^\mu(\tau) A_\mu(x(\tau)) \right]\,,
\label{eq:wloop}
\eeq
computed on a closed contour $\C_n$. In what follows, the closed contour
is a light-like $n$-edged polygonal contour~\cite{Alday:2007hr}.
The contour is such that labelling the $n$ vertices of the polygon as $x_1,\ldots,x_n$, the distance 
between any two contiguous vertices, {\em i.e.}, the length of the edge in between, is given by the
momentum of a particle in the corresponding colour-ordered scattering amplitude,
\beq
p_i = x_i - x_{i+1}\, ,\label{eq:dist}
\eeq
with $i=1,\ldots,n$. Because the $n$ momenta add up to zero, $\sum_{i=1}^n p_i = 0$, the $n$-edged contour
closes, provided we make the identification $x_1 = x_{n+1}$.

In the weak-coupling limit, the Wilson loop can be computed as an expansion in the coupling.
The expansion of \Eqn{eq:wloop} is done through
the non-abelian exponentiation theorem~\cite{Gatheral:1983cz,Frenkel:1984pz},
which gives the vacuum expectation value of the Wilson loop as an exponential,
\beq
\langle W[\C_n] \rangle = 1 + \sum_{L=1}^\infty a^L W_n^{(L)} = {\rm exp} \sum_{L=1}^\infty a^L w_n^{(L)}\, ,\label{eq:fgt}
\eeq
where the coupling is defined as
\beq
a = \frac{g^2 N}{8\pi^2}\,.
\eeq
For the first two loop orders, one obtains
\beq
w_n^{(1)} = W_n^{(1)}\,, \qquad
w_n^{(2)} = W_n^{(2)} - \frac{1}{2} \left( W_n^{(1)}\right)^2\, .\label{eq:twowl}
\eeq
The one-loop coefficient $w_n^{(1)}$ was evaluated in Refs.~\cite{Drummond:2007aua,Brandhuber:2007yx},
where it was given in terms of the one-loop $n$-point MHV amplitude,
\beq
w_n^{(1)} = \frac{\Gamma(1-2\eps)}{\Gamma^2(1-\eps)} m_n^{(1)} =
m_n^{(1)} - n \frac{\zeta_2}{2} + \ord(\eps) 
\, ,\label{eq:wlamp}
\eeq
where the amplitude is a sum of one-loop {\em two-mass-easy} box functions~\cite{Bern:1994zx},
\beq
m_n^{(1)} = \sum_{p,q} F^{\rm 2m e}(p,q,P,Q)\,, \label{eq:2me}
\eeq
where $p$ and $q$ are two external momenta corresponding to two opposite massless legs, 
while the two remaining legs $P$ and $Q$ are massive. The two-loop coefficient $w_n^{(2)}$
has been computed analytically for $n=4$~\cite{Drummond:2007cf} and $n=5$~\cite{Drummond:2007au} and
numerically for $n=6$~\cite{Drummond:2008aq} and $n=7, 8$~\cite{Anastasiou:2009kna}.

In Ref.~\cite{Drummond:2007au} it was established that 
the Wilson loop fulfils a special conformal Ward identity, whose solution is the ansatz plus, for $n\ge 6$, 
an arbitrary function of the conformally invariant cross-ratios, defined in \Eqn{eq:crossratios}.
Thus, the two-loop coefficient $w_n^{(2)}$ can be written as
\beq
w_n^{(2)}(\eps) = f^{(2)}_{WL}(\eps)\, w_n^{(1)}(2\eps) + C_{WL}^{(2)} + R_{n,WL}^{(2)} + \ord(\eps)\, ,\label{eq:wl2wi}
\eeq
where the constant is the same as in \Eqn{eq:ite2bds}, $C_{WL}^{(2)} = C^{(2)} = -\zeta_2^2/2$, 
and the function $f^{(2)}_{WL}(\eps)$ is~\cite{Drummond:2007cf,Anastasiou:2009kna,Korchemskaya:1992je}
%
\beq
f^{(2)}_{WL}(\eps) = -\zeta_2 + 7\zeta_3\eps - 5\zeta_4\eps^2\, .\label{eq:fdue}
\eeq
With the two-loop coefficient $w_n^{(2)}$ given by \eqns{eq:wl2wi}{eq:fdue} and the two-loop MHV amplitude given by
\eqns{eq:ite2bds}{eq:discr}, the duality between Wilson loops and amplitudes is expressed by the equality of their
remainder functions~\cite{Drummond:2008aq,Anastasiou:2009kna},
\beq
R_{n,WL}^{(2)} = R_n^{(2)}\,. \label{eq:remaind}
\eeq

Defining the conformally invariant cross ratios as,
\beq
u_{ij} = \frac{x_{ij+1}^2 x_{i+1j}^2}{x_{ij}^2x_{i+1j+1}^2}\, ,\label{eq:crossratios}
\eeq
for $n=6$, they are~\cite{Drummond:2007au},
\beq
u_{36} = u_1 = \frac{x_{13}^2 x_{46}^2}{x_{36}^2 x_{41}^2}\,, \qquad 
u_{14} = u_2 = \frac{x_{15}^2 x_{24}^2}{x_{14}^2 x_{25}^2}\,, \qquad 
u_{25} = u_3 = \frac{x_{26}^2 x_{35}^2}{x_{25}^2 x_{36}^2}\,, \label{eq:3cross}
\eeq
where $x_{ij}^2 = (x_i-x_j)^2$, and using \Eqn{eq:dist} one sees that $x_{i,i+2}^2 = s_{i,i+1}$ and $x_{i,i+3}^2 = s_{i,i+1,i+2}$,
where the labels are understood to be modulo 6.

\section{The quasi-multi-Regge kinematics of a cluster along the ladder}
\label{sec:qmrk}

As we remarked in the Introduction, it suffices to compute $R_{6,WL}^{(2)}$ in any kinematical limit which 
does not modify the analytic dependence of $R_{6,WL}^{(2)}$ on $u_1,u_2,u_3$.
The simplest of those limits to feature an exact Regge factorisation of $w_6^{(2)}$, and in fact in general
of $w_6^{(L)}$,
is the QMRK of a pair along the ladder~\cite{Fadin:1989kf,DelDuca:1995ki}.
In those kinematics, the outgoing gluons are strongly ordered in rapidity, except for a central pair of gluons along the ladder,
while their transverse momenta are all of the same size. In the physical region,
defining $1$ and $2$ as the incoming gluons and $3, 4, 5, 6$ as the outgoing gluons, the ordering can be chosen as
\begin{equation}
y_3 \gg y_4 \simeq y_5 \gg y_6;\qquad |p_{3\perp}| \simeq |p_{4\perp}| 
\simeq |p_{5\perp}| \simeq|p_{6\perp}|\, 
,\label{qmrk6ptc}
\end{equation}
where the particle momentum $p$ is parametrised in terms of the rapidity $y$ and the azimuthal angle $\phi$,
$p=(|p_{\perp}|\cosh y, |p_{\perp}|\cos\phi, |p_{\perp}|\sin\phi,|p_{\perp}|\sinh y)$.
We shall work in the Euclidean region, where the Wilson loop is real.
There the Mandelstam invariants are taken as all negative, and in the QMRK of a pair along the ladder 
they are ordered as follows,
\begin{equation}
-s_{12} \gg -s_{34}, -s_{56}, -s_{345}, -s_{123} \gg -s_{23}, -s_{45}, -s_{61}, -s_{234}\, .\label{eq:mrk2lLip} 
\end{equation}
Introducing a parameter $\lambda \ll 1$, the hierarchy above is equivalent to the rescaling
\beq
\{ s_{34}, s_{56}, s_{123}, s_{345} \} = \ord(\lambda)\,, \qquad
\{ s_{23}, s_{45}, s_{61}, s_{234} \} = \ord(\lambda^2)\,.  \label{eq:qmrla}
\eeq
It is easy to see that in this limit the three conformally invariant cross-ratios (\ref{eq:3cross}) do not take trivial limiting 
values~\cite{DelDuca:2008jg},
\beq\bsp
u_1&\,\rightarrow u_1^{QMRK}  = \frac{s_{45}}{(p_4^++p_5^+)(p_4^-+p_5^-)} = \ord(1)\, , \\
u_2&\,\rightarrow u_2^{QMRK}= \frac{|p_{3\perp}|^2  p_5^+p_6^-}{(|p_{3\perp}+p_{4\perp}|^2 + p_5^+p_4^-) 
(p_4^++p_5^+)p_6^- } = \ord(1)\, ,\\
u_3&\,\rightarrow u_3^{QMRK} = \frac{|p_{6\perp}|^2  p_3^+p_4^- }{p_3^+ (p_4^-+p_5^-) 
(|p_{3\perp}+p_{4\perp}|^2 + p_5^+p_4^-) } = \ord(1)\, .\label{thrinvarqmrkc}
\esp\eeq

A similar analysis can be carried through for the seven-edged Wilson loop, $w_7^{(L)}$.
We have verified that the simplest limit to feature an exact Regge factorisation
is the QMRK of three-of-a-kind along the ladder~\cite{Del Duca:1999ha}.
In the physical region, the outgoing gluons are strongly 
ordered in rapidity, except for a cluster of three along the ladder,
\begin{equation}
y_3 \gg y_4 \simeq y_5 \simeq y_6 \gg y_7 ;\qquad |p_{3\perp}| \simeq |p_{4\perp}| 
\simeq |p_{5\perp}| \simeq|p_{6\perp}| \simeq|p_{7\perp}|\, 
.\label{qmrk7trec}
\end{equation}
In the Euclidean region, the Mandelstam invariants are ordered as follows,
\beq\bsp
-s_{12} &\,\gg  -s_{123}, -s_{345}, -s_{567}, -s_{712}, -s_{34}, -s_{67} \gg \\
&\,\gg -s_{23}, -s_{45},  -s_{56}, -s_{71}, -s_{234}, -s_{456}, -s_{671}\, .\label{eq:qmrk7treceucl} 
\esp\eeq
Through a parameter $\lambda \ll 1$, the hierarchy above is equivalent to the rescaling
\beq\bsp
\{ s_{123}, s_{345}, s_{567}, s_{712}, s_{34}, s_{67} \} &\,= \ord(\lambda)\,,\\
\{ s_{23}, s_{45}, s_{56}, s_{71}, s_{234}, s_{456}, s_{671} \} &\,= \ord(\lambda^2)\,. \label{eq:7scale2trec}
\esp\eeq
Using \Eqn{eq:crossratios}, and the fact that $x_{ij+1}^2 = s_{i\cdots j}$, it is easy to see that the seven cross ratios
of the seven-edged Wilson loop do not take trivial limiting values under the rescaling (\ref{eq:7scale2trec}),
\beq
\{u_{14}, u_{25}, u_{36}, u_{47}, u_{51}, u_{62}, u_{73}\} = \ord(1) \,.\label{eq:7cross}
\eeq
Thus, the dependence of $w_7^{(L)}$ on the seven cross ratios is not modified by the QMRK
of three-of-a-kind along the ladder (\ref{eq:qmrk7treceucl}), and hence $w_7^{(L)}$ undergoes
an exact Regge factorisation in this limit.

The same pattern unfolds for the eight-edged Wilson loop, $w_8^{(L)}$.
The simplest limit to feature an exact Regge factorisation 
is the QMRK of four-of-a-kind along the ladder~\cite{claude}.
In the physical region, the outgoing gluons are strongly 
ordered in rapidity, except for a cluster of four along the ladder,
\begin{equation}
y_3 \gg y_4 \simeq y_5 \simeq y_6 \simeq y_7 \gg y_8; \qquad |p_{3\perp}| \simeq |p_{4\perp}| 
\simeq |p_{5\perp}| \simeq|p_{6\perp}| \simeq|p_{7\perp}| \simeq|p_{8\perp}|\, 
.\label{qmrk8trec}
\end{equation}
In the Euclidean region, the Mandelstam invariants are ordered as follows,
\beq\bsp
-s_{12} &\,\gg -s_{1234}, -s_{3456}, -s_{123}, -s_{345}, -s_{678}, -s_{812}, -s_{34}, -s_{78} \gg \\
&\,\gg -s_{2345}, -s_{4567}, -s_{234}, -s_{456}, -s_{567}, -s_{781},
-s_{23}, -s_{45},  -s_{56}, -s_{67}, -s_{81}. \label{eq:qmrk8quadceucl} 
\esp\eeq
Through the parameter $\lambda \ll 1$, the hierarchy above corresponds to the rescaling
\beq\bsp
\{ s_{1234}, s_{3456}, s_{123}, s_{345}, s_{678}, s_{812}, s_{34}, s_{78} \} &\,= \ord(\lambda)\, ,\\
\{ s_{2345}, s_{4567}, s_{234}, s_{456}, s_{567}, s_{781}, s_{23}, s_{45}, s_{56}, s_{67}, s_{81} \} 
&\,= \ord(\lambda^2)\,. \label{eq:8scale}
\esp\eeq
It is easy to check that all the twelve cross ratios of the eight-edged Wilson loop
do not take trivial limiting values under the rescaling (\ref{eq:8scale}),
\beq
\{u_{14}, u_{25}, u_{36}, u_{47}, u_{58}, u_{61}, u_{72}, u_{83}, u_{15}, u_{26}, u_{37}, u_{48}\}
= \ord(1) \,.\label{eq:8cross}
\eeq
Thus, the dependence of $w_8^{(L)}$ on the twelve cross ratios is not modified by the QMRK
of four-of-a-kind along the ladder (\ref{eq:qmrk8quadceucl}), and hence $w_8^{(L)}$ undergoes
an exact Regge factorisation in this limit.

The pattern above generalises to the $n$-edged Wilson loop, $w_n^{(L)}$.
We illustrate briefly how the QMRK of a cluster of 
$(n-4)$-of-a-kind along the ladder features the exact Regge factorisation of $w_n^{(L)}$.
In the physical region, the outgoing gluons are strongly 
ordered in rapidity, except for a cluster of $(n-4)$ along the ladder, 
\begin{equation}
y_3 \gg y_4 \simeq \ldots \simeq y_{n-1} \gg y_n; \qquad |p_{3\perp}| \simeq \ldots
\simeq|p_{n\perp}|\, .\label{qmrkmc}
\end{equation}
In order to display the exact Regge factorisation of $w_n^{(L)}$, we deal with the cases of an even,
$n=2r$, and an odd, $n=2r+1$, number of edges separately.
Through a parameter $\lambda \ll 1$, the hierarchy of the Mandelstam invariants implied by \Eqn{qmrkmc}
can be rendered by requiring that $s_{12}=\ord(1)$ and by the rescaling
\beq\bsp
s_{12\ldots j} &\,= \ord(\lambda)\,, \qquad\qquad 3\le j\le r\,,\\
s_{34\ldots j} &\,= \ord(\lambda)\,, \qquad\qquad 4\le j\le r+2\, ,\label{eq:qmrkm}
\esp\eeq
for any $n$ and $r\ge 3$. In addition,
\beq\bsp
s_{j\ldots 2r} &\,= \ord(\lambda)\,, \qquad\qquad r+2\le j\le 2r-1\,, \qquad r\ge 3\,,\\
s_{j\ldots 12} &\,= \ord(\lambda)\,, \qquad\qquad r+4\le j\le 2r\,, \qquad\qquad r\ge 4
\, ,\label{eq:qmrkmeven}
\esp\eeq
for $n=2r$, with the labels understood to be modulo $2r$, and
\beq\bsp
s_{j\ldots 2r+1} &\,= \ord(\lambda)\,, \qquad\qquad r+2\le j\le 2r\,,\\
s_{j\ldots 12} &\,= \ord(\lambda)\,, \qquad\qquad r+4\le j\le 2r+1\,, \label{eq:qmrkmodd}
\esp\eeq
for $n=2r+1$ and $r\ge 3$, with the labels understood to be modulo $2r+1$.
All other invariants rescale to be $\ord(\lambda^2)$. It is easy to check that for $n=6, 7, 8$,
\eqnss{eq:qmrkm}{eq:qmrkmodd} reproduce the rescaling of Eqs.~(\ref{eq:qmrla}),
(\ref{eq:7scale2trec}) and (\ref{eq:8scale}).

In order to compare with
the rescaling of Eqs.~(\ref{eq:qmrla}), (\ref{eq:7scale2trec}) and (\ref{eq:8scale}), it is convenient
to classify the cross ratios as follows,
\beq\bsp
u_{1j} &\,= \frac{x_{1j+1}^2 x_{2j}^2}{x_{1j}^2x_{2j+1}^2} =
\frac{s_{1\cdots j}s_{2\cdots j-1}}{s_{1\cdots j-1}s_{2\cdots j}}\,, \qquad\qquad j=4,\ldots, r+1\,,\\
u_{2j} &\,= \frac{x_{2j+1}^2 x_{3j}^2}{x_{2j}^2x_{3j+1}^2} =
\frac{s_{2\cdots j}s_{3\cdots j-1}}{s_{2\cdots j-1}s_{3\cdots j}}\,, \qquad\qquad j=5,\ldots, r+2\,,\\
&\,\,\,\,\vdots\\
u_{2r+1j} &\,= \frac{x_{2r+1j+1}^2 x_{1j}^2}{x_{2r+1j}^2x_{1j+1}^2} =
\frac{s_{2r+1\cdots j}s_{1\cdots j-1}}{s_{2r+1\cdots j-1}s_{1\cdots j}}\,, \qquad j=3,\ldots, r\,,
\label{eq:crossratiosodd}
\esp\eeq
for $n=2r+1$, and
\beq\bsp
u_{1j} &\,= \frac{x_{1j+1}^2 x_{2j}^2}{x_{1j}^2x_{2j+1}^2} =
\frac{s_{1\cdots j}s_{2\cdots j-1}}{s_{1\cdots j-1}s_{2\cdots j}}\,, \qquad\qquad\qquad j=4,\ldots, r+1\,,\\
&\,\,\,\,\vdots\\
u_{rj} &\,= \frac{x_{rj+1}^2 x_{r+1j}^2}{x_{rj}^2x_{r+1j+1}^2} =
\frac{s_{r\cdots j}s_{r+1\cdots j-1}}{s_{r\cdots j-1}s_{r+1\cdots j}}\,, \qquad\qquad j=r+3,\ldots, 2r\,,\\
u_{r+1j} &\,= \frac{x_{r+1j+1}^2 x_{r+2j}^2}{x_{r+1j}^2x_{r+2j+1}^2} =
\frac{s_{r+1\cdots j}s_{r+2\cdots j-1}}{s_{r+1\cdots j-1}s_{r+2\cdots j}}\,, \qquad j=r+4,\ldots, 2r\,,\\
&\,\,\,\, \vdots\\
u_{2rj} &\,= \frac{x_{2rj+1}^2 x_{1j}^2}{x_{2rj}^2x_{1j+1}^2} =
\frac{s_{2r\cdots j}s_{1\cdots j-1}}{s_{2r\cdots j-1}s_{1\cdots j}}\,,\qquad\qquad\quad j=3,\ldots, r-1\,,\\
\label{eq:crossratioseven}
\esp\eeq
for $n=2r$, where we use the fact that $u_{i,i+1}=u_{i,i+2}=0$.
Without further imposing the Gram-determinant constraints,
the counting yields $n(n-5)/2$ conformally invariant cross ratios, in agreement with Ref.~\cite{Anastasiou:2009kna}.
It is easy to check that for $n=6, 7, 8$, \eqns{eq:crossratiosodd}{eq:crossratioseven} generate
the cross ratios of Eqs.~(\ref{eq:3cross}), (\ref{eq:7cross}) and (\ref{eq:8cross}). 
Then by inspection
one can see that the cross ratios (\ref{eq:crossratiosodd}) and (\ref{eq:crossratioseven}) do not take limiting 
values under the rescaling of Eqs.~(\ref{eq:qmrla}), (\ref{eq:7scale2trec}) and (\ref{eq:8scale}).
Thus, the dependence of $w_n^{(L)}$ on the cross ratios (\ref{eq:crossratiosodd}) and 
(\ref{eq:crossratioseven}) is not modified by the QMRK of a cluster of $(n-4)$-of-a-kind along the
 ladder (\ref{qmrkmc}), and hence $w_n^{(L)}$ undergoes
an exact Regge factorisation in this limit.
 
 Finally, we note that the exact Regge factorisation of the $n$-edged Wilson loops, with $n= 4, 5$,
 may be dealt with as a degenerate case of the QMRK of a cluster of $(n-4)$-of-a-kind along the ladder.
 Namely, for $n= 4$ one obtains the QMRK of a cluster of zero particles along the ladder, which is the standard 
 Regge limit~\cite{Drummond:2007aua}, and for $n= 5$ the QMRK of a cluster of one particle along the ladder,
which is the multi-Regge kinematics of five particles~\cite{Bartels:2008ce,Brower:2008nm}.

\section{The two-loop six-edged Wilson loop}
\label{sec:w62}

In Ref.~\cite{Anastasiou:2009kna} an expression for a generic two-loop $n$-edged Wilson loop as a sum of Euler-type integrals was presented, similar to Feynman parameter integrals appearing in the computation of Feynman integrals. In Sec.~\ref{sec:qmrk} we argued that the
Wilson loops are Regge exact in the QMRK where $(n-4)$ gluons are emitted along the ladder. Hence, it follows that it is sufficient to compute the individual integrals in the QMRK to obtain the Wilson loop in \emph{general} kinematics.

In the present work we concentrate exclusively on the two-loop six-edged Wilson loop, $w_6^{(2)}$, which is the first case where the remainder function is non zero. Hence, an analytic computation of $w_6^{(2)}$ is equivalent to an analytic computation of the two-loop six-point remainder function $R_{WL,6}^{(2)}$. 
We start from the parametric representations for the diagrammatic
contributions to the two-loop $n$-edged Wilson loop derived in Ref.~\cite{Anastasiou:2009kna},
and we derive appropriate Mellin-Barnes representations  for them using the standard formula,
\begin{equation}
{1\over (A+B)^\lambda} = {1\over \Gamma(\lambda)}\,{1\over 2\pi i}\int_{-i\infty}^{+i\infty}\rd z\,\Gamma(-z)\,\Gamma(\lambda+z)\,{A^{z}\over B^{\lambda+z}}\, ,
\end{equation}
where the contour is chosen such as to separate the poles in $\Gamma(\ldots-z)$ from the poles in $\Gamma(\ldots+z)$.
Note that in our case $\lambda$ is in general an integer plus an off-set corresponding to the dimensional regulator $\eps$. In order to resolve the singularity structures in $\epsilon$, we apply the
strategy based on the Mellin-Barnes representation and given in Refs.~\cite{Smirnov:1999gc,Tausk:1999vh,Smirnov:2004ym,Smirnov:2006ry}. To this effect, we apply the codes {\tt MB}~\cite{Czakon:2005rk} and {\tt MBresolve}~\cite{Smirnov:2009up} and obtain a set of Mellin-Barnes integrals that can be safely expanded in $\eps$ under the integration sign. Then we proceed to take the QMRK limit defined by Eq.~(\ref{eq:mrk2lLip}) by applying {\tt MBasymptotics} \cite{CzakonMBA} to extract the leading QMRK behavior of each Mellin-Barnes integral, and {\tt barnesroutines} \cite{Kosower} to perform
integrations that can be done by corollaries of Barnes lemmas. 
To be more explicit, in Ref.~\cite{Anastasiou:2009kna} the six-edged Wilson loop was expressed as,
\beq
w_6^{(2)}  =  \begin{cal}C\end{cal}\,\sum_i\,f_i(p_k)\,,
\eeq
where $f_i(p_k)$ denote the parametric integrals of Ref.~\cite{Anastasiou:2009kna} depending on the external momenta $p_k$. The prefactor $\begin{cal}C\end{cal}$ is defined by
\beq
\begin{cal}C\end{cal} = 2a^2\mu^{4\eps}[\Gamma(1+\eps)\,e^{\gamma_E\eps}]^2\,.
\eeq
and the scale $\mu^2$ is given in terms of the Wilson loop scale, $\mu^2_{WL} = \pi\,e^{\gamma_E\eps}\mu^2$. The Regge exactness of $w_6^{(2)}$ enables us to write
\beq\label{eq:wllim1}
w_6^{(2)}  =  \begin{cal}C\end{cal}\,\sum_i\,f_i^{(1)}(p_k)\,,
\eeq
where $f_i^{(1)}(p_k)$ denotes the leading asymptotic behavior of $f_i(p_k)$ in the QMRK limit defined by the scaling~(\ref{eq:qmrla}),
\beq
f_i(p_k) = f_i^{(1)}(p_k) + \ord(\lambda)\,.
\eeq
In Section~\ref{sec:qmrk} we considered the QMRK limit where gluons 1 and 2 are incoming, \emph{i.e.}, where $s_{12}$ is the largest invariant. Of course there are five additional ways in which we could have defined the limit, corresponding to the cyclic permutations of the external gluons, \emph{e.g.}, we could have considered the QMRK limit defined by the scaling,
\beq
\{ s_{45}, s_{61}, s_{234}, s_{123} \} = \ord(\lambda)\,, \qquad
\{ s_{34}, s_{56}, s_{12}, s_{345} \} = \ord(\lambda^2)\,,  \label{eq:qmrla2}
\eeq
where $\lambda\ll1$ and $s_{23} = \ord(1)$. Note that this limit is incompatible with the limit~(\ref{eq:qmrla}), \emph{i.e.}, terms that are $\ord(\lambda)$ in one limit can be large in another limit, and vice-versa. However, the Regge exactness of the Wilson loop allows us to iterate this procedure and to repeat the previous argument starting from \Eqn{eq:wllim1} and to take the limit~(\ref{eq:qmrla2}). Then we arrive at
\beq\label{eq:wllim2}
w_6^{(2)}  =  \begin{cal}C\end{cal}\,\sum_i\,f_i^{(2)}(p_k)\,,
\eeq
where $f_i^{(2)}(p_k)$ is the leading asymptotic behavior of $f_i^{(1)}(p_k)$ in the limit~(\ref{eq:qmrla2}),
\beq
f_i^{(1)}(p_k) = f_i^{(2)}(p_k) + \ord(\lambda)\,.
\eeq
 It is straightforward to see how this procedure iterates for the remaining four cyclic permutations of \Eqn{eq:qmrla}.
Finally, we arrive at a set of multiple Mellin-Barnes integrals $f^{(6)}_i(p_k)$ of a much simpler type than the original ones. After applying our procedure, all integrals are at most threefold and all of them are explicitly dependent on the conformal cross ratios only, because the cross ratios are the only combination of invariants that are invariant under all the cyclic permutations of \Eqn{eq:qmrla}. However, note that the coefficients of the integrals do not depend only on the conformal cross ratios, but also on logarithms of Mandelstam invariants, which arise when expanding the MB integrals in a QMRK limit of the type defined in \Eqn{eq:qmrla}, as can be easily seen from the following example,
\beq
{1\over 2\pi i}\int_{-1/2-i\infty}^{-1/2+i\infty}\rd z\,\Gamma(-z)\,\Gamma(z)\,\lambda^z\,\left({s_{12}\over s_{23}}\right)^z = \ln{s_{12}\over s_{23}} + \ord(\lambda)\,.
\eeq
Finally, we checked numerically that the sum of the Mellin-Barnes integrals in the QMRK, $f^{(6)}_i(p_k)$, is equal to the sum of all the original parametric integrals, the latter being evaluated numerically using {\tt FIESTA}~\cite{Smirnov:2008py, Smirnov:2009pb}.

The resulting Mellin-Barnes integrals are then evaluated by directly closing contours and summing up residues or by exchanging a Mellin-Barnes integration with an integral of Euler type. The infinite sums which appear in the intermediate steps of the computation are typically generalised harmonic sums~\cite{Moch:2001zr, Moch:2005uc} as well as multiple binomial sums~\cite{Jegerlehner:2002em, Kalmykov:2007dk}. The convergence of the series requires the conformal cross ratios to be less than 1, and in the following we concentrate on this kinematic region, within the Euclidean region. Details on the explicit computation of the integrals will be presented in a forthcoming 
publication~\cite{us}. 
Here it suffices to say that, except for the contribution coming from the hard diagram with six light-like edges, all the integrals can be expressed in terms of harmonic polylogarithms~\cite{Remiddi:1999ew} in one conformal cross ratio. In turn, the six-edged hard diagram constitutes the bulk
of the final result, and 
can be written as a linear combination of Goncharov's multiple polylogarithms~\cite{Goncharov:1998}, whose arguments are functions of conformal cross ratios. These polylogarithms are defined by the iterated integration,
\beq
G(\vec w; z) = \int_0^z{\rd t\over t-a}\,G(\vec w';t) {\rm ~~and~~} G(\vec 0_n;z) = {1\over n!}\ln^nz\, ,
\eeq
where we define $\vec w = (a, \vec w')$, and for $z=1$ they are manifestly real, if all the elements in the weight vector $\vec w$ are either greater than 1 or negative. The number of elements of $\vec w$ is called the (transcendental) weight of $G(\vec w;z)$. The polylogarithms we obtain can be divided into several classes, corresponding to the elements $w_i$ of the weight vector,
\begin{enumerate}
\item $w_i=1/u_j, 1/(1-u_j), (1-u_j)/(1-u_j-u_k)$. 

It is easy to see that in this case $w_i>1$ or $w_i<0$, for $0<u_i,u_j<1$.
\item $w_i=1/(u_i+u_j)$. 

In this case $w_i$ could be smaller than 1, \emph{i.e.}, the polylogarithms can develop an imaginary part. However, we checked numerically that the imaginary parts cancel in the final answer.
\item $w_i=1/u_{jkl}^{(\pm)}, 1/v_{jkl}^{(\pm)}$, where we define
\beq\bsp\label{eq:sqrt}
u_{jkl}^{(\pm)}=&\,\frac{1-u_j-u_k+u_l\pm\sqrt{\left(u_j+u_k-u_l-1\right)^2-4 \left(1-u_j\right)\left(1-u_k\right) u_l}}{2 \left(1-u_j\right) u_l}\, ,\\
v_{jkl}^{(\pm)} =&\,\frac{u_k-u_l\pm\sqrt{-4 u_j u_k u_l+2 u_k u_l+u_k^2+u_l^2}}{2  \left(1-u_j \right)u_k}\, .
   \esp
\eeq
\end{enumerate}
A comment is in order about the square roots in Eq.~(\ref{eq:sqrt}). It turns out that the square roots become complex for certain values of the conformal cross ratios inside the unit cube, but they always come in pairs such that the sum of the two contributions is real. To emphasize this property, we introduce the following notation,
\beq\bsp
\begin{cal}G\end{cal}(\ldots, u_{ijk},\ldots; z) &\, = G\left(\ldots,u_{ijk}^{(+)},\ldots;z\right) +G\left(\ldots,u_{ijk}^{(-)},\ldots;z\right)\, ,\\
\begin{cal}H\end{cal}(\vec w;1/u_{ijk}) &\,= H \left(\vec w;1/u_{ijk}^{(+)}\right) +H \left(\vec w;1/u_{ijk}^{(-)}\right)\, ,
\esp\eeq
and similarly for $v_{ijk}^{(\pm)}$.

After having computed the contributions from the individual integrals, we can easily extract the remainder function $R_{WL,6}^{(2)}$ by subtracting the contribution from our computation, thus obtaining the first fully analytic representation of $R_{WL,6}^{(2)}$ in the
Euclidean region. Note that although we performed the computation in the QMRK of a pair along the ladder, as introduced in Sec.~\ref{sec:qmrk}, the Regge exactness of the Wilson loop in this limit ensures that our expression is valid in general kinematics. The final result for the remainder function can be expressed as a linear combination of multiple polylogarithms of uniform transcendental weight four\footnote{In the present version of the remainder function, we had to extend the definition of the transcendental weight to include the imaginary roots, which are present in \Eqn{eq:sqrt}.
However, given that the remainder function is real, we cannot exclude that in a suitable basis those imaginary roots could disappear.}. 
Because the result is rather lengthy, we present it in an electronic form at {\tt www.arxiv.org} where a text file containing the {\tt Mathematica} expression for the remainder function is provided. 

We have checked numerically that our result is completely symmetric in its arguments. Furthermore, we have checked analytically that the expression satisfies the constraints imposed by the multi-Regge and the collinear limits. Note that the vanishing in these limits is non trivial, since the expression of $R_{WL,6}^{(2)}$ in general kinematics involves polylogarithms whose arguments are ratios of cross ratios, which can be $\ord(1)$ in the limit. However, all those contributions exactly cancel when approaching the limit. Finally, we have checked numerically at several points that our results agree with the numerical results of Ref.~\cite{Anastasiou:2009kna}. 

\begin{center}
\begin{figure}[!t]
\begin{center}
\includegraphics{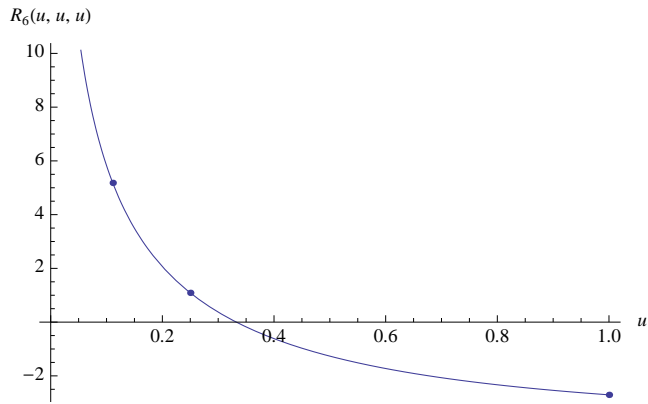}
\caption{\label{fig:plot}The remainder function $R_6^{(2)}(u,u,u)$ for $0< u\le1$. The points represent the numerical values given in Ref.~\cite{Anastasiou:2009kna}.}
\end{center}
\end{figure}
\end{center}

In the particular case where all three conformal cross ratios are equal, we find that,
\beq\bsp
R_{WL,6}^{(2)}(1,1,1) &\,= -{\pi^4\over 36} \simeq -2.70581...\, ,\\
\lim_{u\to \infty}\,R_{WL,6}^{(2)}(u,u,u) &\, = -{\pi^4\over 144}\simeq -0.67645...\,,
\esp\eeq
in agreement, within numerical errors, with the values quoted in Ref.~\cite{Anastasiou:2009kna}. Similarly, the asymptotic behavior for $u\to0$ is given by,
\beq
\lim_{u\to0}\,R_{WL,6}^{(2)}(u,u,u) = {\pi^2\over 8}\ln^2u + {17\pi^4\over 1440} + \ord(u)\,.
\eeq
 Further results for the special case where all three conformal ratios are equal are summarized in Fig.~\ref{fig:plot}. Note that even though our numerical evaluation is for the moment limited to $0<u_i\le 1$, we can still compute the asymptotic value when all conformal cross ratios are equal and large by expanding the Mellin-Barnes integrals around $u=\infty$ \emph{before} taking residues. We find perfect agreement with the numerical value quoted in Ref.~\cite{Anastasiou:2009kna}, which deviates slightly from the asymptotic value obtained from the analytic expression of the remainder function proposed in Ref.~\cite{Alday:2009dv}.

We conclude this section by making some comments on the numerical evaluation of Goncharov's multiple polylogarithms. Up to weight two, Goncharov's polylogarithms can be expressed in terms of ordinary logarithms and dilogarithms, \emph{e.g.}\footnote{Note that these expressions are valid for generic values of the parameters $a$ and $b$. For the limiting cases where the parameters approach 0 or 1 some care is needed, \emph{e.g.}, $G(0;z) = \ln z$, whereas $\lim_{a\to 0}\ln\left(1-{z\over a}\right)$ is divergent.},
\beq\bsp
G(a;z) &\,= \ln\left(1-{z\over a}\right)\, ,\\
G(a,b;z) &\,=
\text{Li}_2\left(\frac{b-z}{b-a}\right)-\text{Li}_2\left(\frac{b}{b-a}\right)+
   \log \left(1-\frac{z}{b}\right) \log \left(\frac{z-a}{b-a}\right)\, .
   \esp\eeq
However, our result involves polylogarithms up to weight four. We observe that in these cases the polylogarithms can be evaluated in a numerically stable and fast way, even for complex arguments, by writing them as an iterated integration of polylogarithms of weight two,
\beq\bsp
G(a,b,c;z) &\, = \int_0^z{\rd t_1\over t_1-a}\,G(b,c;t_1),\\
G(a,b,c,d;z) & \,= \int_0^z{\rd t_1\over t_1-a}\,\int_0^{t_1}{\rd t_2\over t_2-b}\,G(c,d;t_2),
\esp\eeq
and using {\tt Mathematica}'s native {\tt NIntegrate}  command to perform the integration, provided that the integrals converge.

\section{Conclusions}

In this letter we have identified a class of kinematics for
which the multi-loop $n$-edged Wilson loop exhibits exact Regge factorisation and which
leave invariant the analytic form of the Wilson loop.
In those kinematics, the analytic result for the Wilson loop is the same as in general kinematics,
although the computation is remarkably simplified with respect to general kinematics. 
Using the simplest of those kinematics,
the QMRK of a pair along the ladder, we have performed the first analytic computation of the 
two-loop six-edged Wilson loop in general kinematics.
The computation has been done in the Euclidean region, where the result is real. 
Except for the contribution coming from the hard diagram with six light-like edges, the result can be expressed in terms of harmonic polylogarithms in one conformal cross ratio. In turn, the six-edged hard diagram, which constitutes the bulk of the final result,
can be written as a linear combination of Goncharov's multiple polylogarithms, whose arguments are functions of conformal cross ratios.
Finally, the remainder function can be expressed as a linear combination of multiple polylogarithms of uniform transcendental weight four.
Details on the explicit computation of the integrals will be presented in a forthcoming publication~\cite{us}. 

\section*{Acknowledgements}

We thank Ugo Aglietti, Babis Anastasiou, Andi Brandhuber, Mikhail Kalmykov,
Gregory Korchemsky and Gabriele Travaglini for useful discussions.
In particular, we thank Nigel Glover for useful discussions and for the fruitful collaboration which led to the present work, Paul Heslop and Valya Khoze for providing the numerical value of the two-loop six-point remainder function at several points,
and Alexander Smirnov for providing a version of FIESTA suitable to compute the Wilson-loop integrals. VDD and CD thank the CERN Theory Group and CD and VAS thank the Laboratori Nazionali di Frascati
for the hospitality during various stages of this work.
This work was partly supported by RFBR, grant 08-02-01451, and
by the EC Marie-Curie Research Training Network ``Tools and Precision
Calculations for Physics Discoveries at Colliders'' under contract MRTN-CT-2006-035505.

\end{document}